\documentclass[12pt]{iopart}

\usepackage[english]{babel}
\usepackage[utf8]{inputenc}
\usepackage{graphicx}
\usepackage{xcolor}
\usepackage{amssymb}
\usepackage{subfigure}
\usepackage{anysize}
\usepackage[author={Max Schlepzig}]{pdfcomment}

\newcommand{\ket}[1]{\ensuremath{\left|{#1}\right\rangle}}

\date{\today}

\providecommand{\newoperator}[2]{\newcommand*{#1}{\mathop{\mathrm{#2}}\nolimits}}
\newoperator{\sgn}{sgn}
\newoperator{\arctanh}{arctanh}
\newoperator{\diag}{diag}

\newcommand{\openone}{\ensuremath{1\!\!1}}

\begin{document}

\title{Bose-Hubbard models with photon pairing in circuit-QED}
\author{Benjam\'in Villalonga Correa}
\address{Instituto de F\'isica Fundamental, IFF-CSIC, Serrano 113-bis, Madrid E-28006, Spain}
\author{Andreas Kurcz}
\address{Instituto de F\'isica Fundamental, IFF-CSIC, Serrano 113-bis, Madrid E-28006, Spain}
\author{ Juan Jos\'e Garc{\'\i}a-Ripoll}
\address{Instituto de F\'isica Fundamental, IFF-CSIC, Serrano 113-bis, Madrid E-28006, Spain}

\begin{abstract}
In this work we study a family of bosonic lattice models that combine an on-site repulsion term with a nearest-neighbor pairing term, $\sum_{i} a^\dagger_i a^\dagger_{i+1} + \mathrm{H.c.}$ Like the original Bose-Hubbard model, the nearest-neighbor term is responsible for the mobility of bosons and it competes with the local interaction, inducing two-mode squeezing. However, unlike a trivial hopping, the counter-rotating terms form pairing cannot be studied with a simple mean-field theory and does not present a quantum phase transition in phase space. Instead, we show that there is a cross-over from a pure insulator to long-range correlations that start up as soon as the two-mode squeezing is switched on. We also show how this model can be naturally implemented using coupled microwave resonators and superconducting qubits.%
\end{abstract}

\submitto{\jpb}
\maketitle

\section{Introduction}
\label{sec:intro}

The physics of lattice bosons is an interesting topic that has experienced several revivals associated to the proposal and realization of new experimental setups. These include, for instance, the study of phase transitions in Josephson-junction arrays~\cite{VanderZant1992,VanderZant1996}, the ground-breaking studies of quantum phase transitions with bosonic atoms in optical lattices~\cite{Greiner2002} and, more recently, the proposals for polariton physics in coupled cavities and photonic systems~\cite{Greentree2006,Hartmann2006a,Angelakis2007}. The previous represents a curious round trip where a very interesting physical model is introduced in the context of Condensed Matter Physics, we learn how to tame it and control it with the tools of Quantum Simulation and the final and most promising platform for studying its physics seems to be, once more, a superconducting setup~\cite{Houck2012a,Underwood2012}.

The use of superconducting quantum circuit setups to implement bosonic lattice models has various advantages over atomic implementations, but the most important one is the access to a wider variety of interactions. Using photons as particles we can easily engineer nonlinearities~\cite{Greentree2006,Hartmann2006a,Angelakis2007}, but we also have access to controlled dissipation~\cite{Marcos2012,Porras2012,Jin2013}, easily customizable geometries~\cite{Underwood2012} and effective dispersion relations~\cite{Zueco2012} and, most important, the possibility of increasing the coupling to a point in which counterrotating interactions become relevant~\cite{Forn-Diaz2010,Niemczyk2010}.

In this work we explore the possibility of engineering exotic interactions in circuit-QED, studying models with photons that do not preserve the number of particles. Our study is centered around a model of coupled cavities with pairing and on-site interaction ($\hbar=1$)
\begin{equation}
  \label{eq:model}
  H = {g \over 2}
  \sum_i \left(e^{{\rm{i}}\phi_{i}} a^\dagger_i a^\dagger_{i+1} + \mathrm{H.c.}\right) 
  + U \sum_i a^\dagger_i a^\dagger_i a_i a_i +\sum_i \omega_i a^\dagger_i a_i.
\end{equation}
In contrast with the original Bose-Hubbard model
\begin{equation}
\label{BHM}
  H =- {t \over 2} \sum_i \left(e^{{\rm{i}}\phi_{i}} a^\dagger_i a_{i+1} + \mathrm{H.c.}\right) + U \sum_i a^\dagger_i a^\dagger_i a_i a_i 
\end{equation}
our problem facilitates the mobility of bosons through a pairing term, $a^\dagger_i a^\dagger_{i+1}$, that does not preserve the number of particles. We will show that the competition between pairing, $g$, and on-site repulsion, $U$, does not have a phase transition associated to it. Instead, for any small value of $g$ we find a cross-over mechanism that establishes long-range entanglement and squeezing in the lattice, through a process that cannot be described by a trivial mean-field theory.

Our work is structured in three sections. In Sect.~\ref{sec:model} we propose a quantum simulation of Eq.~(\ref{eq:model}) using an array of superconducting microwave resonators which are coupled through periodically driven SQUIDs~\cite{Peropadre2013}. We will consider important issues, such as state preparation and tomography of the correlations that we estimate in the next two sections. Here we study this model in two different regimes. In Sect.~\ref{sec:linear} we adopt the regime $U=0$ of non-interacting bosons and derive the squeezing and entanglement properties of the resulting array of linear cavities. We will show that the ground state in momentum space is a collection of paired photons with different momenta. The paired momenta can be controlled using the phase $\phi_{i}$ in Eq.~(\ref{eq:model}) and the whole system approaches criticality when $g \simeq \omega$. The competition between this long-range entanglement and the local interaction is the topic of Sect.~\ref{sec:interaction}. There we will add a nonzero value of $U$ and study the problem using the methods of infinite Matrix Product States or iTEBD~\cite{Orus2008}. We will show that now $U$ prevents a breakdown of the model for large $g$ and that it also suppresses entanglement. However, unlike in the Bose-Hubbard model, the transition from the insulator regime to the squeezed multimode system does not happen through a phase transition but through a cross-over. Finally, in Sect.~\ref{sec:conclusions} we summarise our results and discuss possible extensions of this work.

\section{Quantum Simulation}
\label{sec:model}

The model that we have studied so far~(\ref{eq:model}) is not likely to appear in nature: the counterrotating terms normally appear together with rotating terms, in a way that these ones are dominating the dynamics. However, the fact that certain interactions are not commonly available does not mean that we cannot study them in a physical setup. Instead, we can find a highly tunable quantum mechanical system and implement such interactions by controlling its dynamics, in what is known as a ``quantum simulation''~[See Ref.~\cite{Cirac2012} and accompanying articles]. In essence, this is the spirit behind all recent proposals about coupled cavities and polariton-type physics~\cite{Greentree2006,Hartmann2006a,Angelakis2007}, and it will be the way we suggest to implement those models.

\begin{figure}
	\centering
	\includegraphics[width=0.45\textwidth]{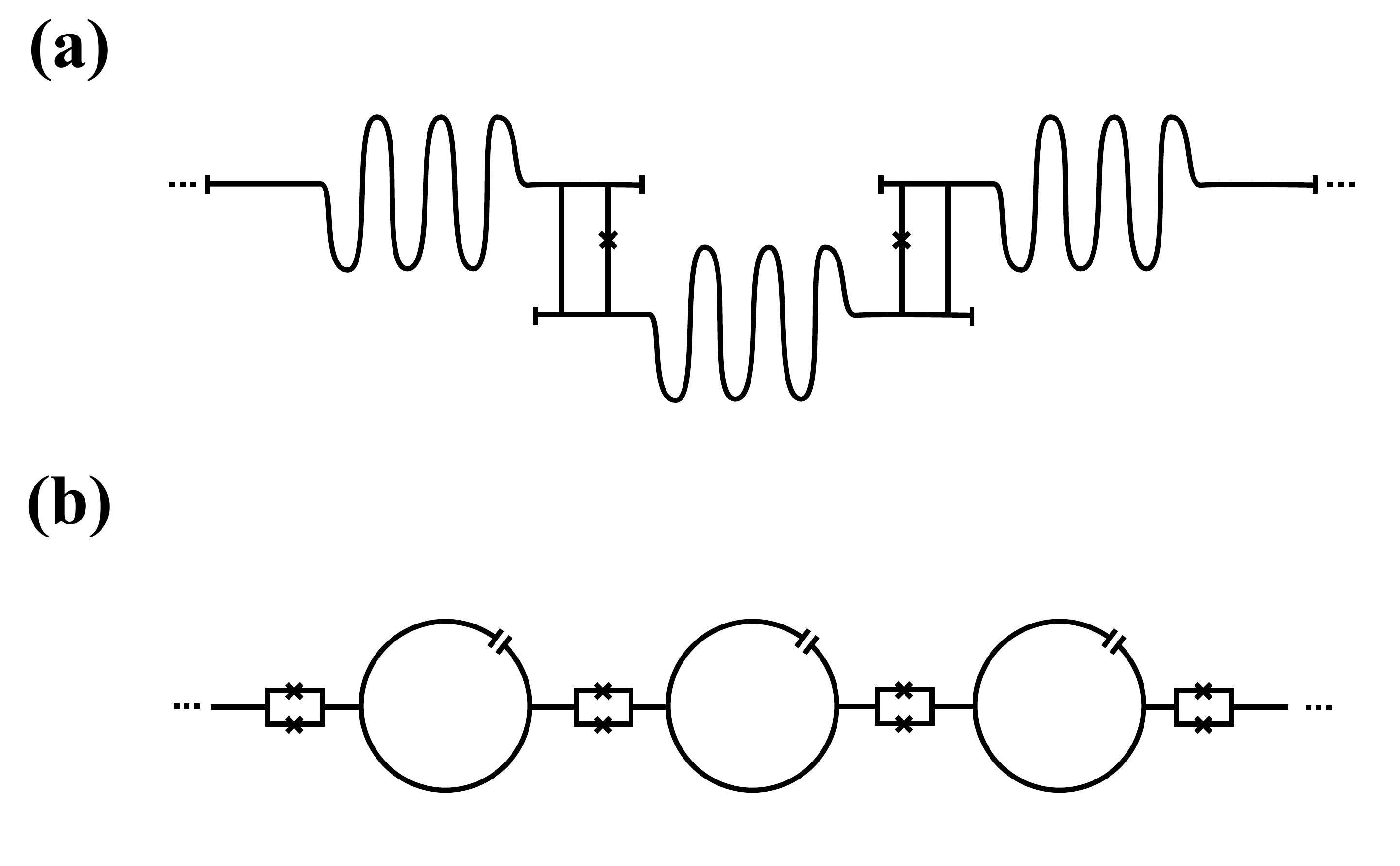}
	\caption{\label{fig:setup}(a) One dimensional array of resonators coupled by means of a superconducting ring coupler. (b) One dimensional array of circular resonators coupled by dc-SQUIDs. Both lattices are folded in a ring in order to keep periodic boundary conditions. Inductive coupling by Josephson Junctions and intersected loops allow to engineer bipartite interaction between nearest neighbors that is governed by counter-rotating terms. }
\end{figure}

For the simulation of our coupled-cavity model we suggest using a one-dimensional setup of coupled microwave resonators. As shown in Fig.~\ref{fig:setup}, such setup consists basically on a number of superconducting segments (coplanar waveguides or striplines) joined by a nonlinear element called SQUID. Each piece of cable supports a number of quantized standing waves, of which we will only focus on the fundamental mode, $\omega_i a^\dagger_i a_i$, with a frequency that will change from cavity to cavity. As shown elsewhere~\cite{Peropadre2013}, by joining the cables through SQUIDs, we actually implement a tunable coupling element %
\begin{equation}
  H = \sum_i \omega_i a^\dagger_i a_i + \sum_i g[\Phi_i(t)]  (a^\dagger_i +a_i)(a^\dagger_{i+1}+a_{i+1}),
\end{equation}
which can be controlled using the magnetic flux $\Phi$ that runs through the Josephson junction loop.

To activate the counter-rotating (rotating) terms in the previous equation, we only need to drive the flux on each cavity periodically~\cite{Peropadre2013}
\begin{equation}
\label{Phi}
\Phi_i (t) \simeq a _{i} + b_i \cos (\nu_i t + \varphi_i)
\end{equation}
with arbitrary $a_i, b_i$ and phase.  If the magnetic flux  $\Phi$ is driven with $\nu_i$ being on resonance with the counter-rotating terms, the propagation of the rotating terms may be  averaged out and vice versa. For this we only need to fabricate cavities with alternating frequencies
\begin{equation}
  \omega_i = \omega_0 + (-1)^{{\rm{i}}} \delta\omega/2,
\end{equation}
and choose a driving $\nu_i= 2\omega_0$ (alternatively chose $\nu_i=\delta\omega (-1)^i$ in order to activate rotating terms).

Note also that in addition to the resonance, we can also control the phase. The time-origin of the driving propagates through a unitary transformation to the effective model that results, creating the full pairing term from Eq.~(\ref{eq:model}).

The last two ingredients in our system are the interaction and the possible external driving. The last one is easy to achieve, because the inductive or capacitive coupling between a cavity and a nearby cable will introduce the possibility of driving photons in and out of the system. The nonlinearity is a bit trickier, as it involves  a nontrivial photon-photon interaction inside the cavity. We expect that such terms could be ported from existing proposals based on nonlinear resonators~\cite{Leib2012}, but a more interesting approach is that of polariton physics~\cite{Greentree2006,Hartmann2006a,Angelakis2007}, where the interaction with a nearby qubit can provide both attractive and repulsive nonlinearities.

The measurement of properties in this setup is also an interesting topic. By nature, the circuit-QED setups tend to be rather closed, as any additional cable or probe may be regarded as a source of decoherence. However, following Ref.~\cite{Quijandria2012}, we envision an alternative that consists on a single transmission line running parallel to our setup of coupled cavities. Ensuring a weak coupling between both systems we can watch in real time the out-coupling of photons from the cavities into the line, thereby probing the frequency- and momentum-dependent correlations that we will discuss in the next section (c.f.~Fig.~\ref{fig:negativity} in Sec.~\ref{sec:entangled_photons}). In addition to this, it would be interesting to probe the transport properties of these models, to see whether the pairing may assist in the mobility of photons, transmitting correlations through the chain, and probe the lack of insulating phases.

\section{Linear Cavities}
\label{sec:linear}

In this section we will study a uniform model of coupled cavities ($\omega_i = \omega$) without on-site interactions ($U = 0$)
\begin{equation}
  \label{eq:model-U0}
  H_0 = {g \over 2}
  \sum_{i} \left(e^{{\rm{i}}\phi_{i}} a^\dagger_i a^\dagger_{i+1} + \mathrm{H.c.}\right)
  + \omega \sum_i a^\dagger_i a_i.
\end{equation}
Just as in the Bose-Hubbard model, our aim is to find the state of the bosons in the limit in which they are free particles. Since this is a quadratic problem in Fock operators, we will use the formalism of Gaussian states, analyzing the entanglement that arises in the system which, as we will see, is not condensed but forms a multimode squeezed state.

In Section~\ref{sec:rlattice} we will first Fourier transform $H_0$ in order to write it in momentum space, where we will see that it can be expressed as the direct sum of two-mode Hamiltonian. Then, in Section~\ref{sec:two_mode} we will be able to diagonalize it and study its stability conditions.  In Section~\ref{sec:boosts} we will analyze the effect of adding a phase to each site that grows linearly with the lattice. In Section~\ref{sec:entangled_photons} we will study entanglement between photonic modes, i.e. between modes in momentum space.

\subsection{The reciprocal lattice}
\label{sec:rlattice}

For simplicity we will assume that the model in Eq.~(\ref{eq:model-U0}) describes a one-dimensional lattice with periodic boundary conditions, and assume $\phi_{i}=0$. Due to the translational invariance of the problem, it makes sense to rewrite the same model in Fourier space. We perform the transformation
\begin{equation}
\label{eq:fourier-transform}
a_i  = \frac{1}{\sqrt{N}} \sum\limits_{k \in I} e^{-{{\rm{i}}} q_k i} b_{q_k}.
\end{equation}
Here the lattice sites run over the indices $i=0,\ldots,N-1$, and $q_k=2\pi k /N\in(-\pi,\pi]$ is the quasimomentum and we choose the quasimomentum index to run over $I=\{-\left[\frac{N}{2}-1\right],\ldots,\left[\frac{N}2\right]\}$ if $N$ is even or over $I=\{-\left[\frac{N-1}{2}\right],\ldots,\left[\frac{N-1}{2}\right]\}$ if it is odd, forming a symmetric Brillouin zone. This unitary transformation transforms $H_0$ to a sum of Hamiltonians, $H_0=\sum_k H_0^{(q_k)}$ with 
\begin{equation}
\label{eq:H-squeezing}
H_0^{(q)} = \omega  (n_q + n_{-q}) + g \cos(q) \left(b_q b_{-q} + \mathrm{H.c.}\right).
\end{equation}
Note how, except for $q_k=0$ and $q_k=\pi$ (if $N$ is even), $H_0^{(q_k)}$ represents the usual two-mode squeezing interaction between bosonic modes. In the following we will analyze the entanglement that results from this coupling, neglecting the point-like singularities at $q_k=0,\pi$, which do not add much to the physics.

\subsection{Two-mode squeezing}
\label{sec:two_mode}

We diagonalize the momentum Hamiltonian~(\ref{eq:H-squeezing}) using the squeezing transformation \cite{Ripka85}
\begin{equation}
\label{S2}
\mathcal{S}_q (\xi_q) = \exp \left [ {1 \over 2} \left ( \xi_q^* b_q b_{-q} - \xi_q b_{q}^\dagger b^\dagger _{-q} \right ) \right ], \qquad \xi_q=r_q e^{{\rm{i}}\varphi_q}
\end{equation}
obtaining an operator
\begin{equation}
\label{diagonalization_hq}
\tilde{H}_0^{(q)}  = \mathcal{S}_q^{\dagger}(\xi_q) H_0^{(q)} \mathcal{S}_q(\xi_q) = \varepsilon_q \left( b_q^{\dagger} b_q + b_{-q}^{\dagger} b_{-q} \right) + \varepsilon_q - \omega \mbox{.}
\end{equation}
The dispersion relation of this problem is
\begin{equation}
\varepsilon_q = \sqrt{\omega^2 - g^2 \cos(q)^2}.
\end{equation}
In order to warrant the stability of the setup we must constrain the photon-photon interaction to be $|g|\leq \omega$, because otherwise there would be no vacuum for this problem.  Doing so, the squeezing parameter is defined as the oscillating function
\begin{equation}
\label{tanh}
\tanh  r_q  =   \frac{\mid g \cos(q) \mid}{\omega}\,
\end{equation}
which is maximum around $q=0$ and $\pm\pi$. Since we have chosen $\phi_{ij} = 0$, the parameter $\xi_q$ is real, and the phase $\varphi_q$ just absorbs its sign.

\subsection{Momentum boosts}
\label{sec:boosts}
In the full model we considered the possibility of having a site-to-site phase difference, $\phi_i$. If this phase is constant, it may be eliminated through a global phase transformation of the bosonic operators. Instead we will consider it to grow linearly through the lattice. Since we wish to preserve translational invariance and periodic boundary conditions, we find that the dependence of the phase must be
\begin{equation}
\label{phi_i}
  \phi_i = \pi \frac{d}{N}(2 i + 1),\quad d,i \in \{0,\ldots, N-1\}.
\end{equation}
We can engineer a unitary gauge transformation that absorbs this phase
\begin{equation}
  a_i = e^{{\rm{i}}\pi d i/N} \tilde{a}_i.
\end{equation}
The original Hamiltonian $H_0$ is expressed now in terms of $\{\tilde{a}_i,\tilde{a}^\dagger_i\}$, but with $\phi_i=0$. The difference is that the quasimomentum operators for this transformed Hamiltonian
\begin{equation}
  \tilde{b}_q = \frac{1}{\sqrt{N}} \sum_i e^{{\rm{i}} q_k i} \tilde{a}_i =
  \frac{1}{\sqrt{N}} \sum_i e^{{\rm{i}} (q_k - \pi d/N) i} a_i,
\end{equation}
correspond to the quasimomenta $q_k - \pi d/N$ from the original model. In other words, the phase $\phi_i = 2 \pi d /N$ corresponds to a momentum boost of amount $-\pi d/N$. Note that to achieve a similar effect in the Bose-Hubbard model we needed a Peierls transformation with uniform phase.
\subsection{Entangled Photon Pairs}
\label{sec:entangled_photons}
Since the Hamiltonian splits into a sum of commuting terms, each of them acting on a separate pair of momenta, we may regard the total density matrix as a tensor product $\rho = \otimes \rho_q$ of Gaussian states, $\rho_q$, for each of the momenta. Each of these density matrices is best characterized by the covariance matrix (CM).

The CM is written in terms of first and second order statistical moments using the field quadrature operators {$x_{q} = ( b_{q} + b_{q}^\dagger ) / \sqrt{2} $} and {$p_{q} = {\rm i} ( b_{q}^\dagger - b_{q} ) / \sqrt{2}$}.  Defining ${\cal Q}^{(q)}= (x_{q}, \, p_{q}, \,  x_{-q}, \,  p_{-q} ) ^{\rm T}$, the CM is the $4\times 4$ matrix that results from the expectation values ${\cal C}_{nm} = {1 \over 2} \langle {\cal Q}_{n} {\cal Q}_{m} + {\cal Q}_{m} {\cal Q}_{n}  \rangle - \langle {\cal Q}_{n} \rangle \langle {\cal Q}_{m} \rangle$. In complete agreement with any quantum characterization of two-mode Gaussian states \cite{Weedbrook2012}, we find a structure of $2\times 2$ blocks
\begin{equation}
\label{CM}
{\cal C} = \left ( \begin{array}{cc}
\alpha & \gamma \\
\gamma^{\rm T} & \beta \\
\end{array} \right )
\end{equation}
where  $\alpha = \beta = {1 \over 2} \cosh r_q \cdot \openone_2$ and $\gamma = \frac{1}{2}\sinh r_q \cdot \openone_2$. Out of these terms, only $\gamma$ encodes correlations between modes.

\begin{figure}[t]
  \includegraphics[width=\textwidth]{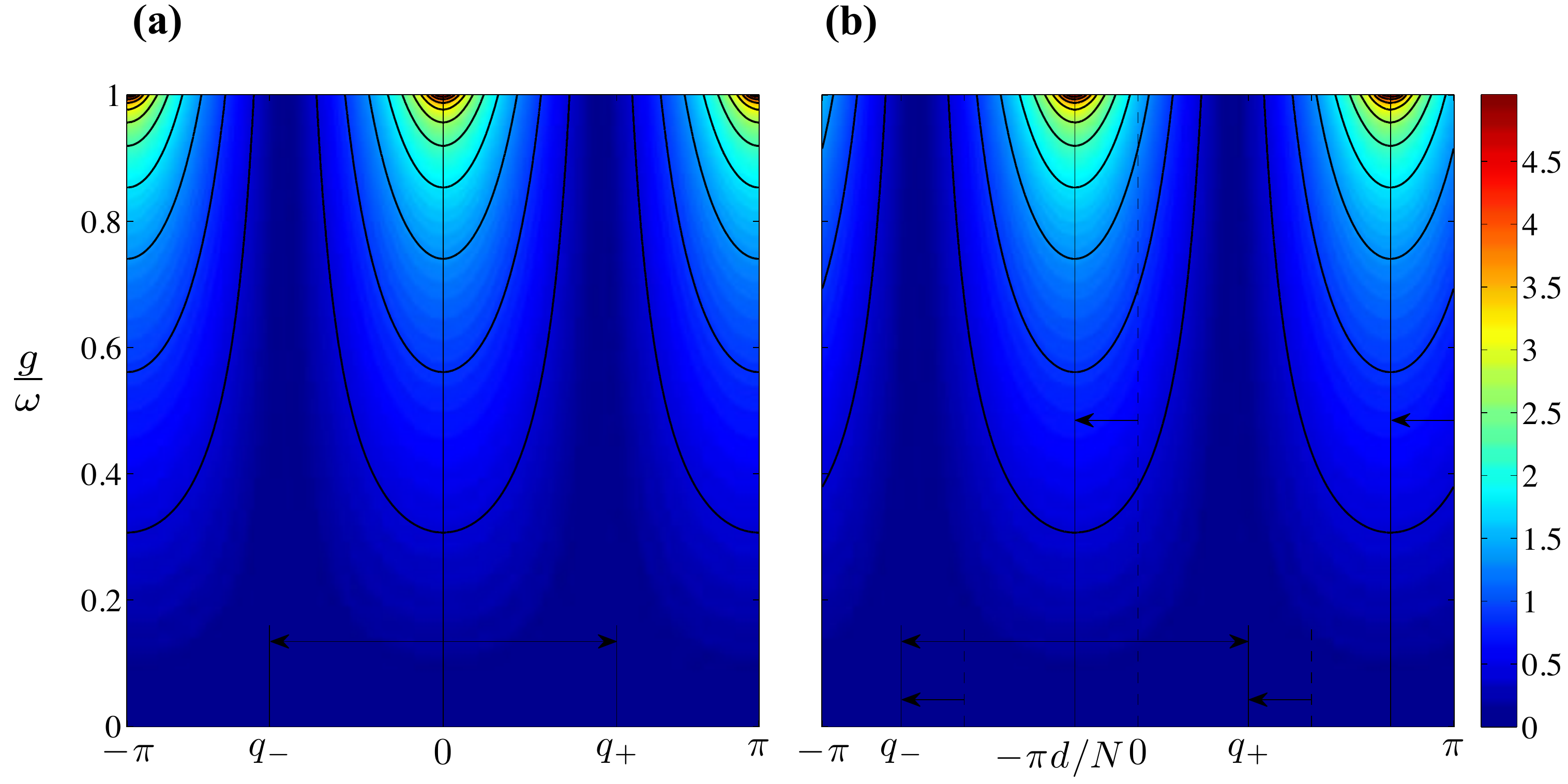}
  \caption{(a) Logarithmic negagtivity $E_{\cal{N}}$ as a function of $g/ \omega$ and \smash{$q$} for $\phi_i=0$. Every quasi momenta $q_-$ is entangled with $q_+=-q_-$. The logarithmic negativity diverges at $q=0$ and $q=\pi$ for $g \rightarrow  \omega$. (b) Similar plot for $\phi_i = \pi \frac{d}{N}(2 i + 1)$ (see Eq.~\ref{phi_i}), with $d=0.2N$. Every quasi momenta $q_-$ is entangled with $q_+=-q_- -2\pi d/N$, and the whole Brillouin zone is boosted by $-\pi d/N$.}
  \label{fig:negativity}
\end{figure}

According to the Peres-Horodecki-Simon criterion \cite{Werner2001,Peres1996, Horodecki1998} the positivity of the partial transpose 
{${\cal{C}}^{\rm{T}_{\rm P}} = \Lambda^{\rm T} {\cal{C}} \Lambda$}, {$\Lambda = \diag{(1,1,1,-1)}$}
represents a sufficient condition for separability of the two-mode state. The entanglement monotone associated to this criterion is the logarithmic negativity \cite{Weedbrook2012, Vidal2002},
\begin{equation}
  E_N = -\frac{1}{2}\sum_i^4 \log_2\left[\min\{1, 2|\nu_i|\}\right],
\end{equation}
where $\nu_i \in \{-\nu_\pm,+\nu_\pm\}$ are the four symplectic eigenvalues of the covariance matrix
\begin{equation}
\label{lpm}
\nu_\pm = \sqrt{{\Delta_1 \pm \sqrt{\Delta_1^2 - 4 \Delta_2} \over 2}} = \frac{1}{2} \sqrt{\frac{1 \pm \tanh r_q }{1 \mp \tanh r_q}},
\end{equation}
where we substituted the invariants $\Delta_1 = \det \alpha + \det \beta - 2 \, \det \gamma$ and $\Delta_2 = \det {\cal C}$ for their value in this particular problem. 

 The immediate consequence is that the negativity is nonzero for any amount of two-mode squeezing
\begin{equation}
\label{logarithmic_negativity_q}
E_{N,q} = \frac{1}{2} \log_2{\left( \frac{1 + | g \cos(q) / \omega |}{1 - | g \cos(q) / \omega |} \right)} .
\end{equation}
In other words, as it was intuitively expected the state is not separable and the entanglement grows with increasing coupling strength, $g$.

In Fig.~\ref{fig:negativity} we show a typical distribution of the entanglement on the first Brillouin zone, both without and with the phase $\phi_i$. In both cases we have pair entanglement between different momenta, which is maximal around to points. The maxima are displaced with the help of the momentum boost $\phi_i$. Using $\phi_i$ we can switch from entangling photons that travel in opposite directions but carry the same energy, to entangling co-propagating photons that differ in frequency, as is the case of $\phi_i\neq 0$ in Fig.~\ref{fig:negativity}b. Note also the similarity between the shape of the entanglement and that of the squeezing (c.~f.~Eq.~(\ref{tanh})).

\section{Pair-Hopping Tunneling}
\label{sec:interaction}

So far we have seen that in the linear case, the counterrotating terms create entanglement and long range correlations. We want to analyze how these correlations compete with an on-site interaction such as the $U$ term from our full model in Eq.~(\ref{eq:model}).

We already know from the physics of the Bose-Hubbard model (cf.~Eq.(\ref{BHM})) a similar competition~\cite{Zwerger2003}: a kinetic term which preserves the number of particles fights for their delocalization against an on-site repulsion that hinders mobility. The result is a quantum phase transition between a superfluid regime, for $t \gg U$, and an insulator regime, $U \gg t$, which has been repeatedly observed using bosonic atoms in optical lattices~\cite{Greiner2002}.

In the model from Eq.~(\ref{eq:model}) the kinetic term is implemented with a pairing term that, by allowing the creation and destruction of pairs of particles in nearby sites, it also allows the mobility of particles. This is obvious from the previous results, the energy bands and the delocalized wavefunctions that result. It is obvious that such a term would compete with the on-site repulsion $U$, but it is not clear under which constraints this competition happens.

A trivial inspection of Eq.~(\ref{eq:model}) reveals that, should we try a similar mean-field treatment as for the Bose-Hubbard, that is replacing $a$ with a nonzero expectation value, $\langle a\rangle$, we would obtain the same phase diagram as for the original model. More precisely, assuming a product state ansatz $\ket{\psi} = \otimes_{i=0}^{N-1} \ket{\phi}$, the energy per site reads
\begin{equation}
  \frac{1}{N} \langle H \rangle _{\omega=0} = \frac{g}{2} \left[\langle a \rangle^2 + \mathrm{H.c.}\right] + U |\langle a \rangle|^4,
\end{equation}
which is minimized for a nonzero expectation value of $a$. Yet we know from the linear model that $\langle a\rangle = 0$ always because of the creation and destruction of pairs of particles. This opens the door to a different type of phase transition, but it also points out the need of a more rigorous and more sophisticated treatment of the problem.

In the following pages we will show that indeed there is no such phase transition, but instead we find a continuous growth of entanglement along the lattice as the pairing $g$ increases and $U$ decreases. This is proven using three different tools: the iTEBD method developed in Ref.~\cite{Vidal2007} and further explained in Ref.~\cite{Orus2008}, DMRG or MPS simulations for the same model~\cite{schollwock2005density}, and perturbation theory in the weak coupling regime, $g \ll U$.

\subsection{Bosonic phase diagram}
\begin{figure}[t]
	\centering
	\includegraphics[width=\textwidth]{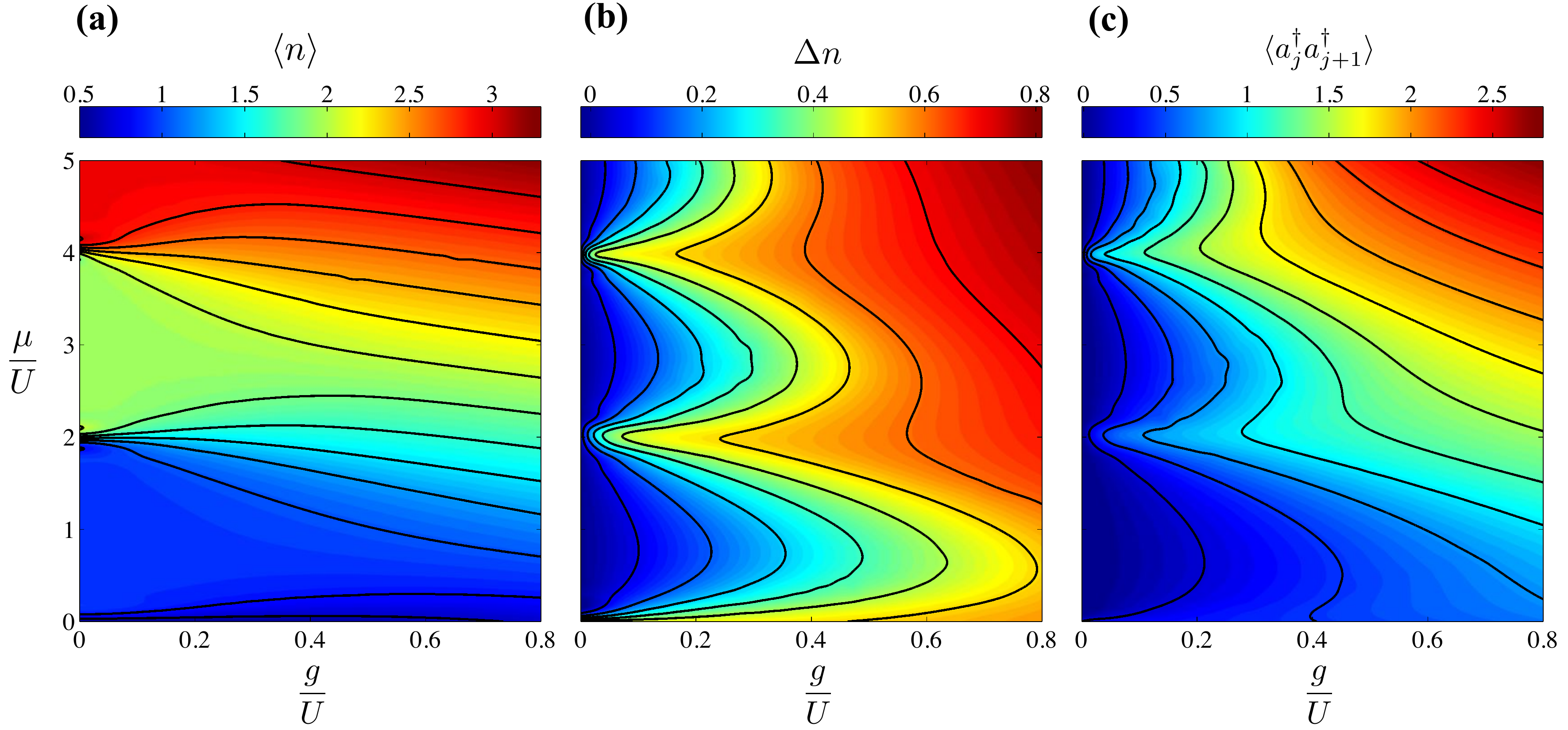}
	\caption{Phase diagram of the free energy $H-\mu N$, for the model in Eq.~(\ref{eq:model}). We plot the (a) number of photons, (b) fluctuations of the photons and (c) the nearest-neighbor pairing term, vs. the chemical potential $\mu$ and the photon-photon pairing $g$. The usual boson delocalization $\langle a^\dagger_i a_{i+1}\rangle$ is zero over the whole diagram. An occupation number cut-off $n_{max} =10$  for the  states $\{i_k\}$ (cf.~Eq.~(\ref{iTEBD})) and $\chi = 20$ has been choosen.}
        \label{fig:phase-diagram}
\end{figure}

The iTEBD method~\cite{Vidal2007,Orus2008} poses a translationally invariant tensor-product state ansatz for a one-dimensional quantum system. Such an ansatz may be qualitatively written as the contraction of four tensors, which are repeated with period two along the lattice,
\begin{equation}
\label{iTEBD}
  \ket{\psi} = \sum \ldots \Gamma^{A,i_0} \Gamma^{B,i_1} \Gamma^{A,i_{2}}
  \ldots \ket{\ldots i_{0},i_1,i_2,\ldots}.
\end{equation}
Here, the $\{i_k\}$ represent the physical states of the $k$-th lattice sites, which in our case is the occupation number of the cavities, cut-off to a reasonable number: $i_k = 0,1,2\ldots n_{max}$. The tensors $\Gamma$ have dimension $\chi^2(n_{max}+1)$, where $\chi$ is another cut-off, this time for the amount of entanglement that the ansatz may host. Simulations have been done ensuring convergence with respect to both cut-offs.

Taking into account the obvious limitations of the ansatz, it is nevertheless possible to get qualitative and even quantitatively accurate answers to many of the physical properties of the ground state. For that we make use of two important properties of Eq.~(\ref{iTEBD}). The first one is that expectation values of local operators, or of products of them (i.e. two- and n-body correlators) can be efficiently computed. The second one is that given an energy functional which is expressed as the sum of local operators, the ansatz above can be optimized to minimize that energy. Both procedures are described in detail in Ref.~\cite{Orus2008} for a detailed explanation of how these quantities are computed through a suitable manipulation and contraction of the tensors in Eq.~(\ref{iTEBD}), where a very efficient algorithm is developed for all these tasks.

We have applied the iTEBD method to studying the properties of the pair hopping model. The study in this section focuses on the ground state of the free energy functional
\begin{equation}
  F = H_{\omega=0} - \mu N,
  \label{eq:free-energy}
\end{equation}
The choice of this functional is motivated by the need of a fair comparison with earlier studies of the Bose-Hubbard model, such as those describing cold atom setups. In this context, the chemical potential $\mu$ is a Lagrange multiplier that enforces an average number of bosons in the system, typically fixed during the loading of atoms in the optical lattice. In the photonic system, enforcing an average number of bosons through an adiabatic preparation is harder to achieve, because the total number of photons in the system is not conserved. It is for this reason that in Sect.~\ref{sec:photonic} we offer an alternative study based on a simpler mechanism to inject photons in the lattice.

\begin{figure}[t]
	\centering
	\includegraphics[width=\textwidth]{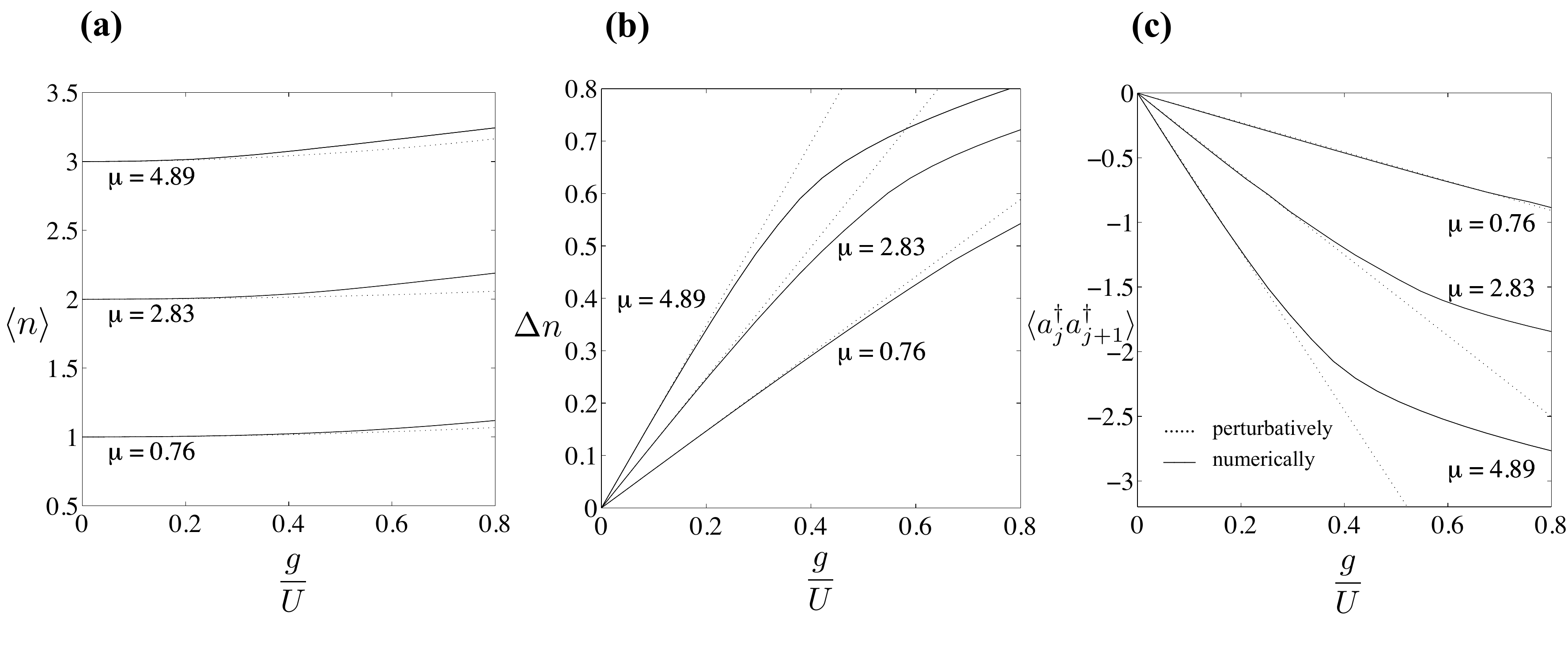}
	\caption{\label{fig:perturbative} Perturbation theory results (dashed) and iTEBD expectation values (solid) for the full model in Eq.~(\ref{eq:model}), close to the insulating phase ($g/U \ll 1$). We plot the (a) number of photons, (b) fluctuations of the photons and (c) the nearest-neighbor pairing term, vs. the photon-photon pairing $g$. The cut-off $n_{max}$ and $\chi$ (cf.~Eq.~(\ref{iTEBD})) are the same as in Fig.~\ref{fig:phase-diagram}.}
\end{figure}

Minimising the mean free energy per site over the set of iTEBD states with fixed bond dimension, $\chi$, we have produced the pictures shown in Fig.~\ref{fig:phase-diagram}, where we plot the density, the number fluctuations $(\Delta n)^2 = \langle (a_i^\dagger a_i) ^2 \rangle - \langle a_i^\dagger a_i \rangle ^2$ and the pairing between nearest neighbors, all for the ground state. In these pictures we appreciate that the lobes of the Bose-Hubbard model are present in the growth of the number fluctuations [cf. Figs.~\ref{fig:phase-diagram}b]. However, the system no longer exhibits plateaus of the density, and there is no sharp phase transition between the insulator and the squeezed or entangled regime. The order parameter of a typical superfluid, which is the largest eigenvalue of the single-particle density matrix, is now zero because  $\langle a^\dagger_i a_{i+1}\rangle = 0$ everywhere. In the paired case, we know that the limit $U\to 0$ is characterized by the two-mode squeezing and long distance correlations. While we do not expect a symmetry breaking even in the presence of interactions, we introduce the pairing observable, $\langle a^\dagger_ia_{i+i}^\dagger \rangle$, as a witness or ``order parameter'' of this squeezed phase. Our simulations confirm that indeed this correlator increases smoothly with the interaction $g/U$, as shown in Fig.~\ref{fig:phase-diagram}c.

The continuity of the transition can be further confirmed by studying weak perturbations around an insulator state. We start with a bare sate that contains exactly $n$ photons per cavity $ | u_n ^0 \rangle = \bigotimes _{i=0} ^{N-1} | n \, \rangle _i$, and apply perturbation theory up to second order, obtaining (cf.~\ref{perturb})
\begin{eqnarray}
\langle a^\dagger_i a_i \rangle &=& n + {g^2 \over 8} \left ( f_{n+1} ^2 - f_n ^2 \right ) + \mathcal{O} \left ( g ^3 \right  ) ~,\\
\langle (a^\dagger_i a_i) ^2 \rangle &=& n^2 +  {g^2 \over 8} \left [ (2 n +1) f_{n+1} ^2 - ( 2n -1) f_n ^2 \right ] +\mathcal{O}  \left ( g ^3 \right  ) ~, \\ 
\langle a_i a_{i+1} \rangle &=& e ^{ {\rm i} \phi_i} {g \over 4} \left [ n f_n- (n+1) f_{n+1} \right ] + \mathcal{O}  \left ( g ^3 \right  ) ~,\\
\langle a_i ^\dagger a_{i+1} \rangle &=&  0 + \mathcal{O}  \left ( g ^3 \right  ) ~,
\end{eqnarray}
expressed in terms of the function $f_n = n\, [2 U (n-1) - \mu]^{-1}$.

Using these expressions we can approximate all observables, including the density, the number fluctuation, etc., and compare with the exact results. This is done in Fig.~\ref{fig:perturbative}, where we find an excellent agreement all the way up to $g=0.1U$, for a variety of filling factors. It is important to remark that perturbation theory confirms the lack of superfluid order, accompanied by the fast growth of the nearest-neighbor squeezing (the quantity that grows fastest of all for small $g$).

\begin{figure}[t]
	\centering
	\includegraphics[width=\textwidth]{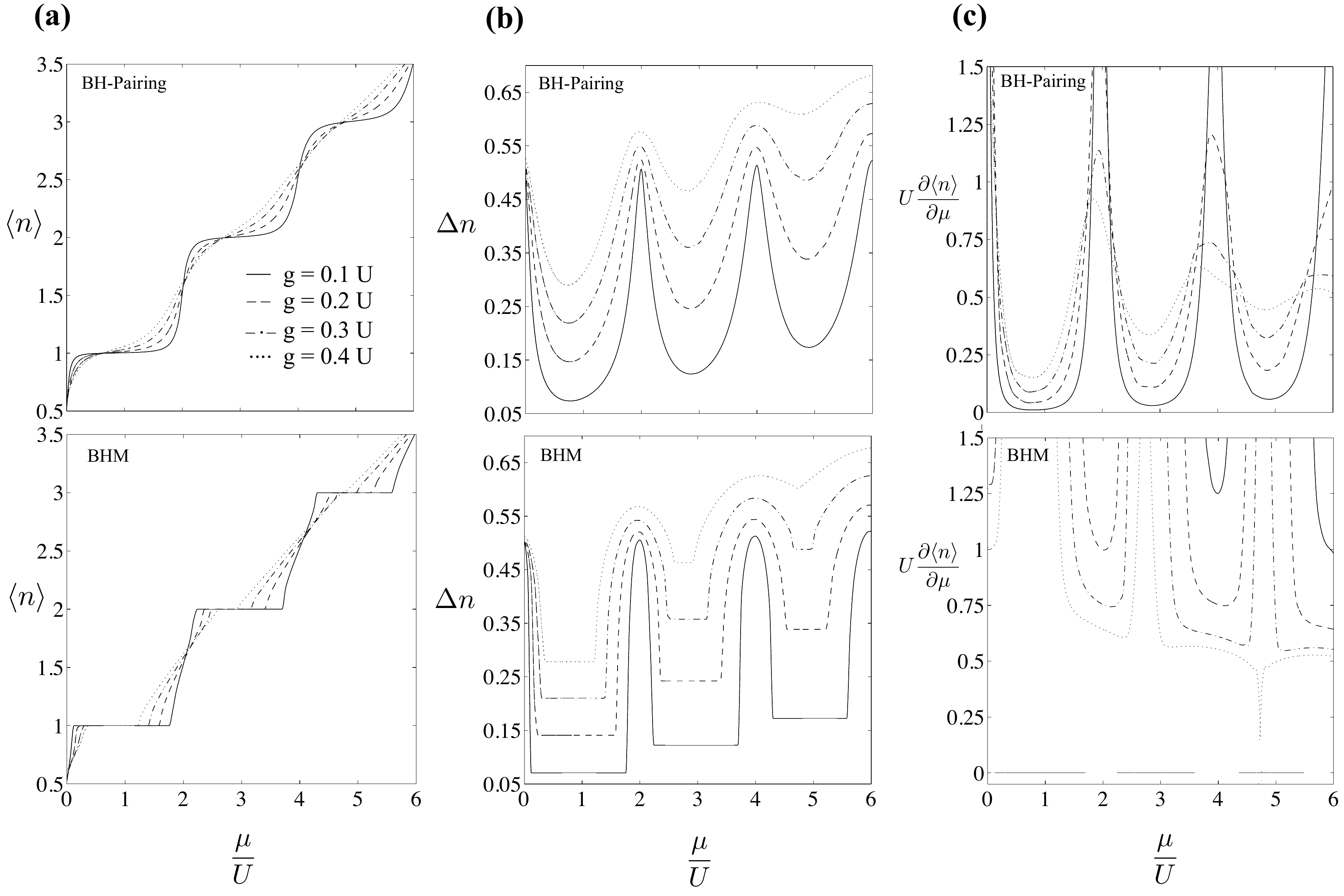}
	\caption{Comparison of the photon pairing model (BH-Pairing) in Eq.~(\ref{eq:model}) and the original Bose Hubbard model (BHM) in Eq.~(\ref{BHM}). We plot for any lattice site $i$ the (a) number of photons, (b) fluctuations of the photons and (c) the compressibility, vs. the chemical potential $\mu$ for four different values of the coupling strength $g$ (see also Fig.~\ref{fig:phase-diagram}).
An occupation number cut-off $n_{max} =10$  for the  states $\{i_k\}$ (cf.~Eq.~(\ref{iTEBD})) and $\chi = 20$ has been choosen. Note how in (c) the compressibility jumps down to zero at the plateaus signaled in (b). Where there are plateaus in (b) there is no compressibility in (c) (the solid line for \smash{$g/U = 0.1$} may cover the lines of vanishing compressibility for the cases \smash{$g/U = 0.2, 0.3, 0.4$}).}
        \label{fig:n_fluc}
\end{figure}

In the previous figures and in the perturbative analysis we have seen that the density and the number fluctuations grow smoothly with the coupling strength and the chemical potential, interpolating from the insulating phase to the squeezed phase. In order to provide further evidence of this continuous behavior, we now add three types of studies: (i) cuts along the phase diagram that show the lack of plateaus, (ii) comparisons with the Bose-Hubbard model to show the differences and the lack of non-analiticities and (iii) further studies involving the long distance correlations in the system, to appreciate whether a phase transition might be hidden in the nonlocal observables.

We start by singling out a fixed photon-photon paring \smash{$g$} and comparing it with the original Bose-Hubbard model in Eq.~(\ref{BHM}). This is done in Fig.~\ref{fig:n_fluc} where we plot the photon number, the photon number fluctuation, and the compressibility $\partial \langle n \rangle /\partial \mu$ for the Bose-Hubbard and the pair hopping models for the same chemical potential $\mu$. In the Bose-Hubbard one can clearly distinguish the superfluid and insulator regions where the photon number $\langle n \rangle$ (number fluctuation $\Delta n$) remains constant and the formation of plateaus becomes evident. The vanishing compressibility within the plateaus confirms the presence of an insulator state followed by a sharp transition into the superfluid regime. This is in  sharp contrast with our model where not even quantities such as the compressibility exhibit sharp jumps other than when approaching the line $g=0$. In our model, the former plateaus of the Bose-Hubbard model now exhibit weak compressibility that is monotonously increasing with the photon-photon pairing $g$. The only valid insulator regime may be observed for the otherwise trivial case $g=0$.

\begin{figure}[t]
	\centering
	\includegraphics[width=\textwidth]{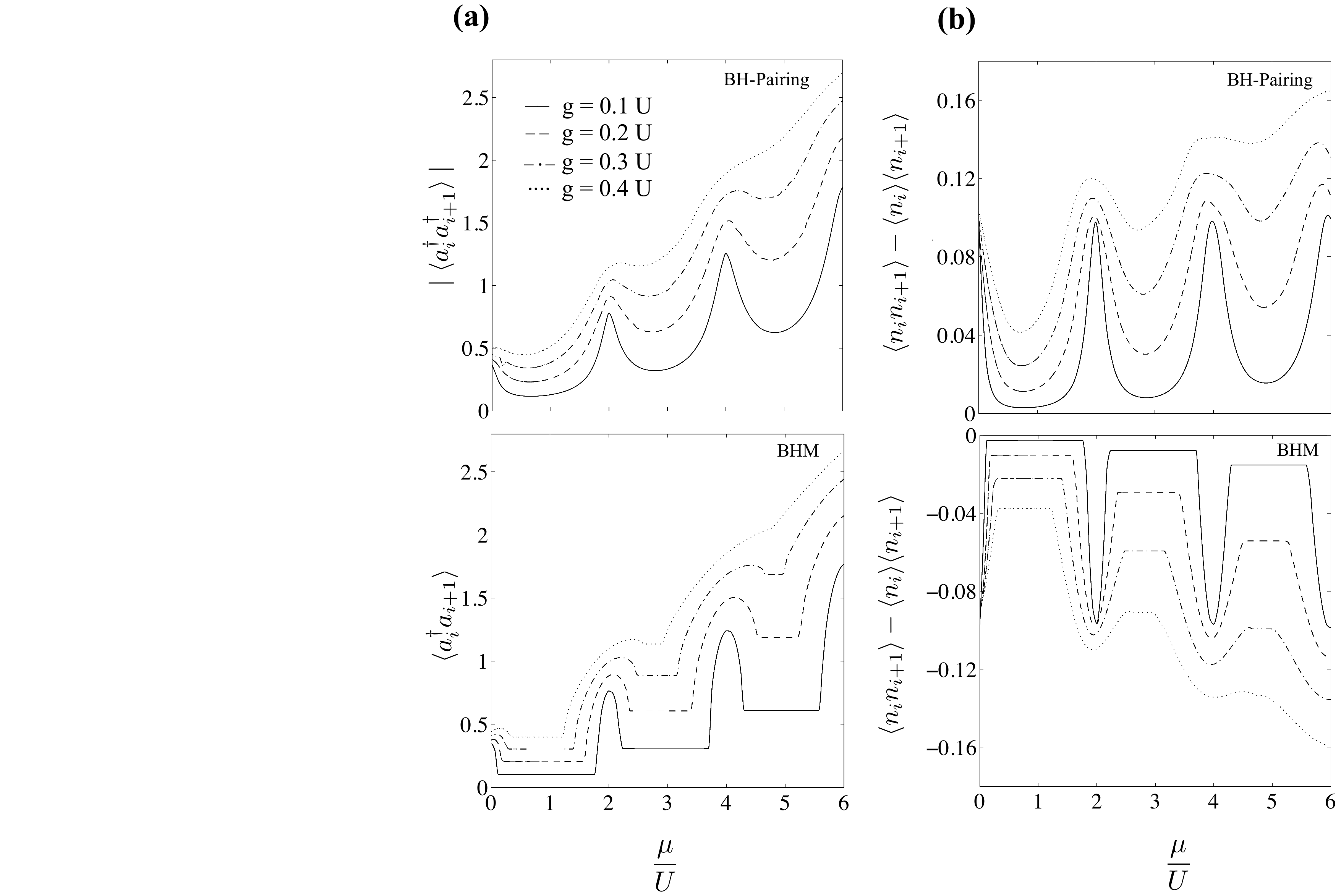}
	\caption{Comparison of the photon pairing model (BH-Pairing) in Eq.~(\ref{eq:model}) and the original Bose Hubbard model (BHM) in Eq.~(\ref{BHM}). We plot the (a) paring term $\langle a^\dagger _i a^\dagger _{i+1} \rangle$ and the boson delocalization $\langle a^\dagger _i a _{i+1} \rangle$, (b) the photon-photon correlation between two neighbouring lattice sites, vs. the chemical potential $\mu$ for four different values of the coupling strength $g$. An occupation number cut-off $n_{max} =10$  for the  states $\{i_k\}$ (cf.~Eq.~(\ref{iTEBD})) and $\chi = 20$ has been choosen.}
        \label{fig:corr}
\end{figure}

We may further stress the lack of plateaus in Fig.~\ref{fig:corr} where we plot the two different order parameters, that is the photon pairing $\langle a^\dagger _i a^\dagger _{i+1} \rangle$ for the model in Eq.~(\ref{eq:model}) and the boson delocalization $\langle a^\dagger _i a _{i+1} \rangle$ for the model in Eq.~(\ref{BHM}), together with the photon-photon correlation \smash{$\langle n_i n_{i+1} \rangle - \langle n_i \rangle \langle n_{i+1} \rangle$} between nearest neighbors of both models. As pointed out before, the crossover in the pair hopping model is clearly driven by squeezing whereas in the Bose-Hubbard model the boson delocalization acts as an adequate order parameter for a superfluid regime\footnote{In all rigor, the actual order parameter would be the largest eigenvalue of the reduced density matrix of the single particle reduced density matrix $\rho_{ij}=\langle a^\dagger_i a_j\rangle$, but since that eigenvalue is associated to a macroscopic population of the zero momentum state, this has as a consequence also the growth of the nearest-neighbor correlator. We expect a similar behavoir for the squeezing.}. The photon-photon correlation is constant in the insulator regime with sharp transitions when the model enters the superfluid regime. In the case of pair-hopping, both the nearest neighbor squeezing and the nearest neighbor density correlations are continously differentiable functions everywhere.
\begin{figure}[t]
	\centering
	\includegraphics[width=\textwidth]{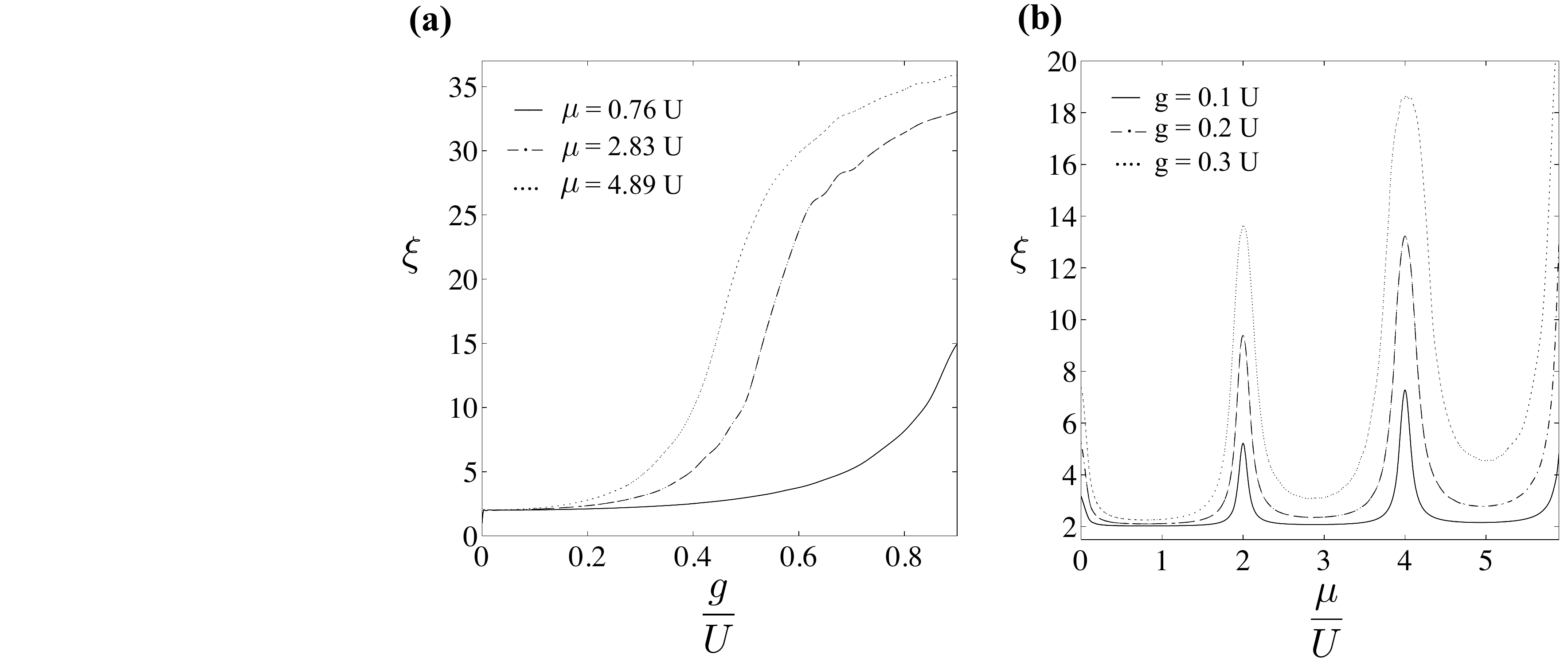}
	\caption{Correlation length $\xi$ for a finite size system of 100 lattice sites folded to a ring for the paring model in Eq.~(\ref{eq:model}). We plot the correlation length for (a) three fixed values of the chemical potential $\mu$ vs. the photon-photon pairing $g$, (b) three fixed values of the photo-photon pairing $g$ vs. the chemical potential $\mu$. An occupation number cut-off $n_{max} =10$  for the  states $\{i_k\}$ (cf.~Eq.~(\ref{iTEBD})) and $\chi = 70$ has been choosen.}
        \label{fig:xi}
\end{figure}

All local observables that we have examined so far show a differentiable behavior typical of a cross-over. However, a phase transition might be hidden in the sudden appearance of long-range order in some observables or correlators. To discard this possibility we have independently studied the correlation length of the squeezing witness $\langle a^\dagger_i a^\dagger_j\rangle$ using finite size simulations with DMRG. It can be shown that in the limit $U\to 0$ this correlator decays algebraically as $1/|i-j|$, with a divergent correlation length. To probe how the photons approach this regime we have computed an average correlation length $\xi (i)$, defined as
\begin{equation}
\xi (i) = { \sum _j \big ( \mid i-j \mid \mid\langle a^\dagger_i a^\dagger_j \rangle - \langle a^\dagger_i \rangle \langle a^\dagger_j \rangle \mid \big ) \over \sum _j \mid\langle a^\dagger_i a^\dagger_j \rangle - \langle a^\dagger_i \rangle \langle a^\dagger_j \rangle \mid} ~,
\end{equation}
for any lattice site $i$. In Fig.~\ref{fig:xi} we show the correlation length $\xi$ for a finite size system of 100 lattice sites. We trace two lines of the phase diagram: a line of fixed chemical potential $\mu$ (Fig.~\ref{fig:xi}a) and that of fixed photon-pairing $g$ (Fig.~\ref{fig:xi}b). In both cases the correlation length grows fast when one leaves the {\sl quasi}-lobes of the bosonic phase diagram in Fig.~\ref{fig:phase-diagram}, but we appreciate no sharp divergence or a boundary. This is most obvious for fixed chemical potential $\mu$. For a fixed photon-paring $g$ the transition among the lobes occurs continuously, with a peak in the region between apparent lobes. This maximum correlation length, however, is never infinite and it grows slowly with $g$.

The previous approach relied on a Condensed Matter Physics phase diagram that combines the interaction with the chemical potential. The chemical potential describes the balance of particles between the system and a reservoir which is put in contact with it. Thus, this approach is very suitable for describing experiments with atoms or electrons, such as lattice models that result from loading atoms in periodic potentials. In a photonic setup, though, we lack superselection rules for the number of photons and indeed these particles are quickly lost after a given time. It is for this reason that instead of achieving equilibrium with a reservoir of photons, it seems more practical to fill the lattice using other mechanisms, such as a coherent drive that injects energy into one or more cavities.

\subsection{Photonic phase diagram}
\label{sec:photonic}
\begin{figure}[t]
	\centering
	\includegraphics[width=\textwidth]{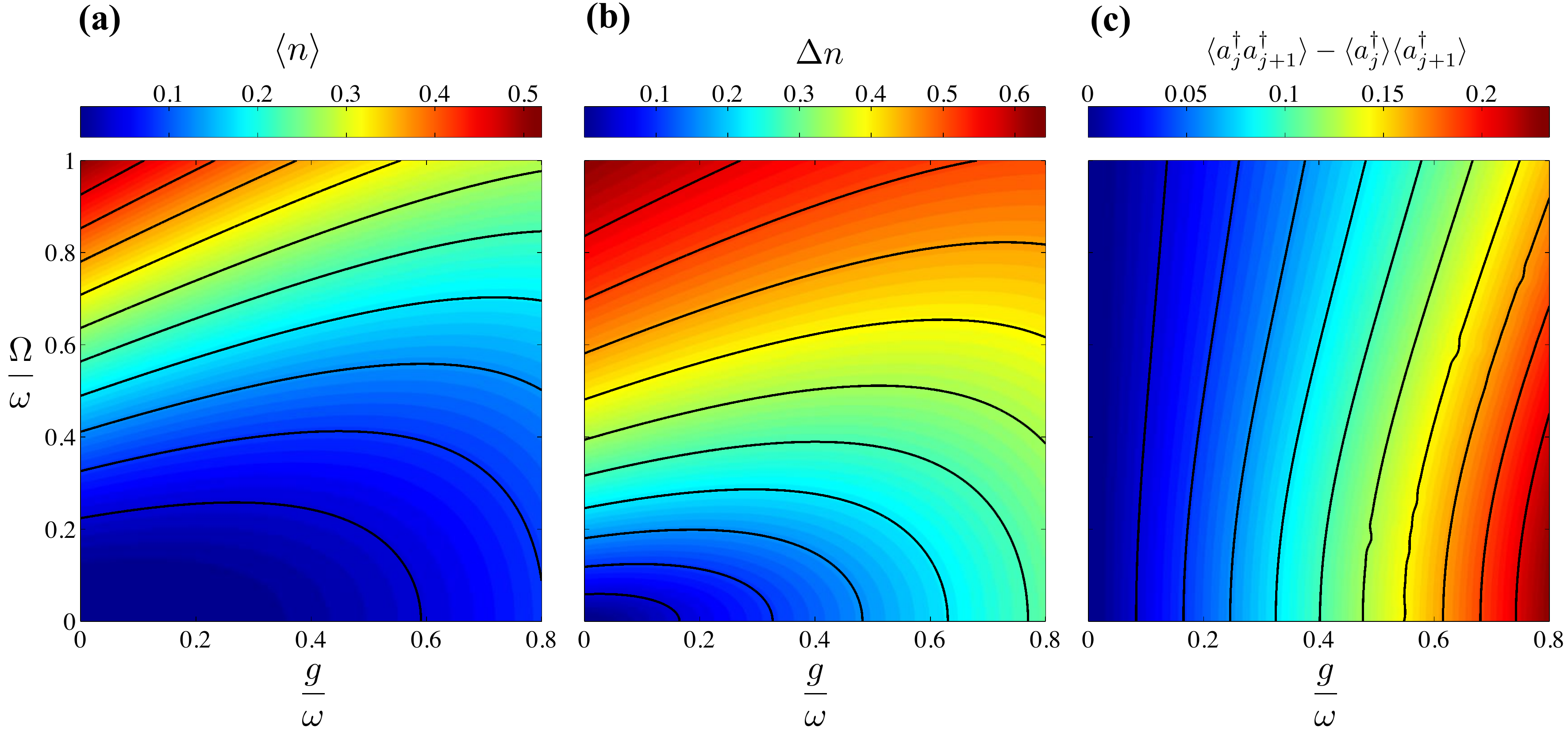}
	\caption{Phase diagram of the driven system $H + \sum_i \Omega(a_i+a_i^\dagger)$, for the model in Eq.~(\ref{eq:driven}). We plot the (a) number of photons, (b) fluctuations of the photons and (c) the nearest-neighbor pairing term, vs. the chemical potential $\mu$ and the photon-photon pairing $g$, taking as units $\omega_i = U = 1$. The usual boson delocalization $\langle a^\dagger_i a_{i+1} \rangle$ is zero over the whole diagram.}
        \label{fig:driven-phases}
\end{figure}

Following this reasoning, and in order to recreate a simpler experiment with superconducting circuits, we have studied an alternative Hamiltonian
\begin{equation}
  H_{\mathrm{driven}} = H + \sum_i \Omega (a_i + a_i^\dagger),
  \label{eq:driven}
\end{equation}
that introduces an external displacement of the cavity. This displacement, of strength $\Omega/\sqrt{\omega}$ in the non-interacting case, is implemented through an external source, which in our case can be a time-independent flux source (i.e. an inductor).

\begin{figure}[t]
	\centering
	\includegraphics[width=\textwidth]{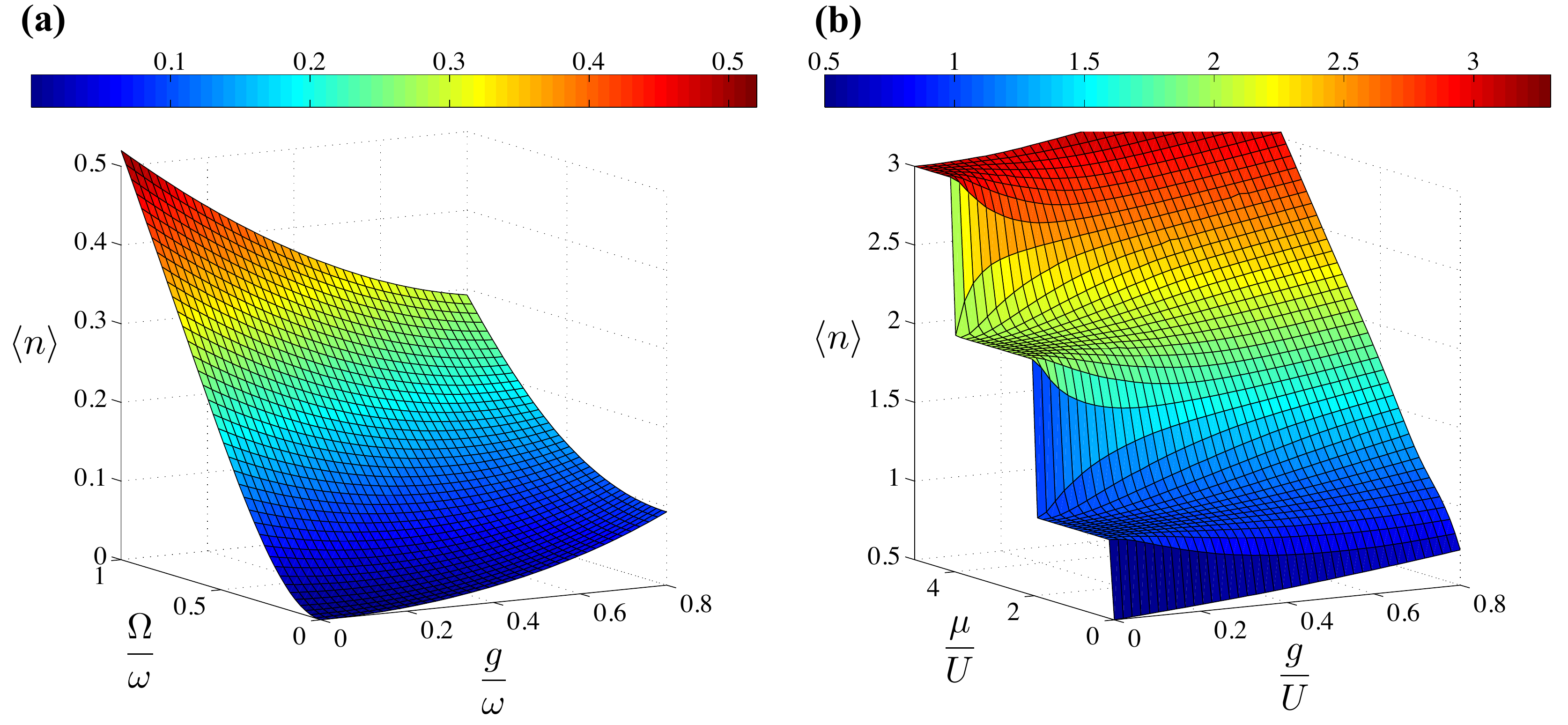}
	\caption{iTEBD expectation values for the number of photons per lattice site. We plot the number of photons for (a) the driven model in Eq.~(\ref{eq:driven}), vs. the photon-photon pairing $g$ and the displacement $\Omega$, (b) the full model in Eq.~(\ref{eq:model}), vs. the photon-photon pairing $g$ and the chemical potential $\mu$. The cut-off $n_{max}$ and $\chi$ (cf.~Eq.~(\ref{iTEBD})) are the same as in Fig.~\ref{fig:phase-diagram}.}
        \label{fig:driven-surf}
\end{figure}

An unfortunate consequence of this formulation is that we lose the lobe structure. For fixed $\omega_i=1, U=1,$ Fig.~\ref{fig:driven-phases} shows the same observables as before (see also Fig.~\ref{fig:driven-surf} for comparison of the model without driving). We appreciate that the only insulating region is now around the vacuum, at $g=\Omega=0$. Pairing continues to be a good order parameter, but it has to be properly defined, eliminating the equilibrium values due to the injection, $\langle a_i^\dagger a^\dagger_{i+1}\rangle - \langle a^\dagger_i\rangle \langle a^\dagger_{i+1}\rangle$. If we do so, then the pairing is relatively insensitive to $\Omega$ and the average number of photons, and grows rapidly with the coupling strength $g$.

\section{Conclusions}
\label{sec:conclusions}

Summing up, in this work we have studied a lattice model of bosons coupled entirely through counter-rotating pairing terms. Unlike its relative the Bose-Hubbard model, the model from Eq.~(\ref{eq:model}) does not experience any quantum phase transition. Instead, for any value of the pairing, correlations are established among cavities, which we quantify using the pairing order parameter $\langle a^\dagger_i a^\dagger_{i+1}\rangle$. 

We have shown that in the non interacting model $U=0$ this pairing is maximal in momentum space, where it leads to the establishment of two-mode squeezing between different momenta. The selection of paired momenta can be controlled through the phase of the driving or the phase in Eq.~(\ref{eq:model}). 

All simulations demonstrate that the introduction of the nonlinear term $U$ transforms our squeezed state, reducing in a continuous way the fluctuations in the number of particles per site, and also the squeezing and the correlation length. The system remains compressible and with a sufficient amount of long-distance entanglement to be of experimental interest.

All these properties, including the numerically computed correlators and two-photon covariance matrix, can be explored in a setup that consists on superconducting resonators coupled by driven SQUIDs. An open problem which escapes the tools and methods shown in this work would be to study the transport properties of these models, to see whether the pairing propagates from the chain to the injected photons.

Finally, we would like to point out that the model that we have studied could also be simulated using ultracold atoms in optical lattices. In this case the pairing term is provided by a Feshbach resonance that splits a molecule into its two fermionic or bosonic constituents. The split atoms would be trapped in an optical lattice which hinders their mobility and provides an on-site interaction $U$. The resulting model has exactly the form of Eq.~(\ref{eq:free-energy}), but demands a more complicated setup than the microwave photonics ideas shown above.

JJG-R and AK acknowledge financial support from the European project PROMISCE, the Spanish MINECO project FIS2012-33022 and the CAM research consortium QUITEMAD S2009-ESP-1594.

\section*{Bibliography}

\bibliographystyle{unsrt}
\bibliography{lib}

\begin{thebibliography}{10}

\bibitem{VanderZant1992}
H.~van~der Zant, F.~Fritschy, W.~Elion, L.~Geerligs, and J.~Mooij.
\newblock Field-induced superconductor-to-insulator transitions in
  josephson-junction arrays.
\newblock {\em Physical Review Letters}, 69(20):2971--2974, November 1992.

\bibitem{VanderZant1996}
H.~van~der Zant, W.~Elion, L.~Geerligs, and J.~Mooij.
\newblock Quantum phase transitions in two dimensions: Experiments in
  josephson-junction arrays.
\newblock {\em Physical Review B}, 54(14):10081--10093, October 1996.

\bibitem{Greiner2002}
M.~Greiner, O.~Mandel, T.~Esslinger, T.~W. H\"{a}nsch, and I.~Bloch.
\newblock Quantum phase transition from a superfluid to a mott insulator in a
  gas of ultracold atoms.
\newblock {\em Nature}, 415(6867):39--44, January 2002.

\bibitem{Greentree2006}
A.~D. Greentree, C.~Tahan, J.~H. Cole, and L.~C.~L. Hollenberg.
\newblock Quantum phase transitions of light.
\newblock {\em Nature Physics}, 2(12):856--861, November 2006.

\bibitem{Hartmann2006a}
M.~J. Hartmann, F.~G. S.~L. Brand\~{a}o, and M.~B. Plenio.
\newblock Strongly interacting polaritons in coupled arrays of cavities.
\newblock {\em Nature Physics}, 2(12):849--855, November 2006.

\bibitem{Angelakis2007}
D.~Angelakis, M.~Santos, and S.~Bose.
\newblock Photon-blockade-induced mott transitions and xy spin models in
  coupled cavity arrays.
\newblock {\em Physical Review A}, 76(3):031805, September 2007.

\bibitem{Houck2012a}
A.~A. Houck, H.~E. T\"{u}reci, and J.~Koch.
\newblock On-chip quantum simulation with superconducting circuits.
\newblock {\em Nature Physics}, 8(4):292--299, April 2012.

\bibitem{Underwood2012}
D.~L. Underwood, W.~E. Shanks, J.~Koch, and A.~A. Houck.
\newblock Low-disorder microwave cavity lattices for quantum simulation with
  photons.
\newblock {\em Physical Review A}, 86(2):023837, August 2012.

\bibitem{Marcos2012}
D.~Marcos, A.~Tomadin, S.~Diehl, and P.~Rabl.
\newblock Photon condensation in circuit quantum electrodynamics by engineered
  dissipation.
\newblock {\em New Journal of Physics}, 14(5):055005, May 2012.

\bibitem{Porras2012}
D.~Porras and J.~J. Garc\'{\i}a-Ripoll.
\newblock Shaping an itinerant quantum field into a multimode squeezed vacuum
  by dissipation.
\newblock {\em Physical Review Letters}, 108(4):043602, January 2012.

\bibitem{Jin2013}
J.~Jin, D.~Rossini, R.~Fazio, M.~Leib, and M.~J. Hartmann.
\newblock Photon solid phases in driven arrays of nonlinearly coupled cavities.
\newblock {\em Physical Review Letters}, 110(16):163605, April 2013.

\bibitem{Zueco2012}
D.~Zueco, J.~J. Mazo, E.~Solano, and J.~J. Garc\'{\i}a-Ripoll.
\newblock Microwave photonics with josephson junction arrays: negative
  refraction index and entanglement through disorder.
\newblock {\em Physical Review B}, 86(2):024503, October 2012.

\bibitem{Forn-Diaz2010}
P.~Forn-D\'{\i}az, J.~Lisenfeld, D.~Marcos, J.~Garc\'{\i}a-Ripoll, E.~Solano,
  C.~Harmans, and J.~Mooij.
\newblock Observation of the bloch-siegert shift in a qubit-oscillator system
  in the ultrastrong coupling regime.
\newblock {\em Physical Review Letters}, 105(23):237001, November 2010.

\bibitem{Niemczyk2010}
T.~Niemczyk, F.~Deppe, H.~Huebl, E.~P. Menzel, F.~Hocke, M.~J. Schwarz, J.~J.
  Garcia-Ripoll, D.~Zueco, T.~H\"{u}mmer, E.~Solano, A.~Marx, and R.~Gross.
\newblock Circuit quantum electrodynamics in the ultrastrong-coupling regime.
\newblock {\em Nature Physics}, 6(10):772--776, July 2010.

\bibitem{Peropadre2013}
B.~Peropadre, D.~Zueco, F.~Wulschner, F.~Deppe, A.~Marx, R.~Gross, and J.~J.
  Garc\'{\i}a-Ripoll.
\newblock Tunable coupling engineering between superconducting resonators: From
  sidebands to effective gauge fields.
\newblock {\em Physical Review B}, 87(13):134504, April 2013.

\bibitem{Orus2008}
R.~Or\'{u}s and G.~Vidal.
\newblock Infinite time-evolving block decimation algorithm beyond unitary
  evolution.
\newblock {\em Physical Review B}, 78(15):155117, October 2008.

\bibitem{Cirac2012}
J.~I. Cirac and P.~Zoller.
\newblock Goals and opportunities in quantum simulation.
\newblock {\em Nature Physics}, 8(4):264--266, April 2012.

\bibitem{Leib2012}
M.~Leib, F.~Deppe, A.~Marx, R.~Gross, and M.~J. Hartmann.
\newblock Networks of nonlinear superconducting transmission line resonators.
\newblock {\em New Journal of Physics}, 14(7):075024, July 2012.

\bibitem{Quijandria2012}
F.~Quijandr\'{\i}a, D.~Porras, J.~J. Garc\'{\i}a-Ripoll, and D.~Zueco.
\newblock Circuit qed bright source for chiral entangled light based on
  dissipation.
\newblock December 2012.

\bibitem{Ripka85}
J.-P. Blaizot and G.~Ripka.
\newblock {\em Quantum Theory of Finite Systems}, volume~1.
\newblock The MIT Press, 2nd edition, 1985.

\bibitem{Weedbrook2012}
C.~Weedbrook, S.~Pirandola, R.~Garc\'{\i}a-Patr\'{o}n, N.~Cerf, T.~Ralph,
  J.~Shapiro, and S.~Lloyd.
\newblock Gaussian quantum information.
\newblock {\em Reviews of Modern Physics}, 84(2):621--669, May 2012.

\bibitem{Werner2001}
R.~F. Werner and M.~M. Wolf.
\newblock Bound entangled gaussian states.
\newblock {\em Physical Review Letters}, 86(16):3658--3661, April 2001.

\bibitem{Peres1996}
A.~Peres.
\newblock Separability criterion for density matrices.
\newblock {\em Physical Review Letters}, 77(8):1413--1415, August 1996.

\bibitem{Horodecki1998}
M.~Horodecki, P.~Horodecki, and R.~Horodecki.
\newblock Mixed-state entanglement and distillation: Is there a “bound”
  entanglement in nature?
\newblock {\em Physical Review Letters}, 80(24):5239--5242, June 1998.

\bibitem{Vidal2002}
G.~Vidal and R.~F. Werner.
\newblock Computable measure of entanglement.
\newblock {\em Physical Review A}, 65(3):032314, February 2002.

\bibitem{Zwerger2003}
W.~Zwerger.
\newblock Mott hubbard transition of cold atoms in optical lattices.
\newblock {\em Journal of Optics B Quantum and Semiclassical Optics},
  5(2):S9--S16, April 2003.

\bibitem{Vidal2007}
G.~Vidal.
\newblock Classical simulation of infinite-size quantum lattice systems in one
  spatial dimension.
\newblock {\em Physical Review Letters}, 98(7):070201, February 2007.

\bibitem{schollwock2005density}
Ulrich Schollw{\"o}ck.
\newblock The density-matrix renormalization group.
\newblock {\em Reviews of Modern Physics}, 77(1):259, 2005.

\bibitem{Cohen77}
C.~Cohen-Tannoudji, B.~Diu, and F.~Lalo\"e.
\newblock {\em Quantum Mechanics}, volume~1.
\newblock Hermann and John Wiley \& Sons. Inc., 2nd edition, 1977.

\end{thebibliography}

\appendix
\section{Pertubation theory up to second order}
\label{perturb}
Motivated by the numerical result, we probe perturbatively the state
\begin{equation}
| u_n ^0 \rangle = \bigotimes _{i=0} ^{N-1} | n \, \rangle _i
\end{equation}
for an equally loaded lattice with uniform population $n$ on each lattice site \smash{$i= 0, \ldots, N-1$}. Considering $N$ lattice sites, the unperturbed hamiltonian, $H_0$, and the perturbative term, $H_1$, are given by
\begin{equation}
\label{H0}
H_0 = U \sum_{i=0}^{N-1} a^\dagger_i a^\dagger_i a_i a_i -\mu \sum_{i=0}^{N-1} a^\dagger_i a_i ~,~~  H_1 = \frac{1}{2}
  \sum_{i=0}^{N-1} \left(e^{{\rm{i}}\phi_{i}} a^\dagger_i a^\dagger_{i+1} + \mathrm{H.c.}\right) ~,
\end{equation}
with \smash{$H = H_0 + g H_1$}. Its unperturbed energy is given by 
\begin{equation}
E_n^0 = N n~\left [u (n-1) -\mu \right ]~.
\end{equation}

The state $|u _n ^0 \rangle$ is not degenerate, hence we may refer to standard Schr\"odinger theory \cite{Cohen77} for the perturbative ansatz. The corresponding expansion in state, 
\smash{$|u _n \rangle = | u_n ^0 \rangle +  g  ~| u_n ^1 \rangle +  g  ^2~| u_n ^2 \rangle + \mathcal{O} (g^3) $}, and energy, \smash{$E_n = E_n^0 + g~E_n^1 + g^2~E_n^2 +  \mathcal{O} (g^3) $}, provides no correction up to fist order in energy, \smash{$E_n^1 \equiv 0$}. This is a direct consequence of the fact that Bloch states are now driven by excitations in pairs. The total number of excitations is no longer preserved.

However, with
\begin{equation}
| u_n ^1 \rangle = \sum _s {\langle  {\rm e} _s | H_{1} | u _n ^0 \rangle \over  E_n ^0 - {\rm e}_s } ~| {\rm e}_s \rangle 
\end{equation}
we may obtain a correction for the eigenstate in first oder in $ g $.
Here, $| {\rm e}_s \rangle$ refers to the rather trivial eigenstates of $H_0$ in Eq.~(\ref{H0}) with eigenenergies ${\rm e}_s$, i.~e. $H_0 | {\rm e}_s \rangle = {\rm e}_s | {\rm e}_s \rangle$.

Consequently the latter provides
\begin{equation} 
\label{u1}
| u_n ^1 \rangle = {f_n \over 4}  \sum _{i=0}^{N-1} e ^{- {\rm i} \phi_i} | \{n-1, n-1\}_i \rangle - {f_{n+1} \over 4}  \sum _{i=0}^{N-1} e ^{ {\rm i} \phi_i} | \{n+1, n+1\}_i  \rangle ~,
\end{equation}
with amplitudes
\begin{equation}
f_n = {n \over {2 U (n-1) - \mu}}~,
\end{equation}
and the notation \smash{$|\{a_1, a_2, \ldots, a_m\}_i \rangle$} marks the position \smash{$i=0, \ldots, N-1$} of the \smash{$m$}-tupel \smash{$a_1, a_2, \ldots, a_m$} in the otherwhise equally loaded state $|u_n^0 \rangle$. For example, taking into account periodic boundary conditions  on the ring, it is \smash{$|\{n-1,n-1\}_{N-1}\rangle = |n-1, n, \ldots, n, n-1\rangle$}.

From Eq.~(\ref{u1}) a second order correction for the energy is immediately found to be
\begin{equation}
E_n^2 = {N \over 8} \left [ n f_n - (n + 1) f_{n+1} \right ] ~.
\end{equation}

Since there is no first order correction in energy, that is \smash{$\langle u_n ^0 | H_1 | u_n ^0 \rangle = 0$}, one may rewrite the second order correction for the state as
\begin{eqnarray}
| u_n ^2 \rangle &=&  \sum _s {\langle {\rm e} _s | H_1 | u _n ^1 \rangle \over E_n ^0 -  {\rm e}_s} ~|  {\rm e}_s \rangle -{N \over 32} \left ( f_n ^2 +  f_{n+1} ^2 \right ) | u_n ^0 \rangle ~. 
\end{eqnarray}
Since we are interested in expectation values that lead to first and second order statistical moments, only the second term in the latter really matters up to second order in $g$. The first term requires at least higher moments in order to have an effect in the same order of $g$. We hence abdicate providing further details.

\end{document}